# A Holographic Diffuser Generalised Optical Differentiation Wavefront Sensor.

L. Marafatto[a], R. Ragazzoni[a], D. Vassallo[a,b], M. Bergomi[a], F. Biondi[a], E. Carolo[a], S. Chinellato[a], M. Dima[a], J. Farinato[a], D. Greggio[a], M. Gullieuszik[a], D. Magrin[a], E. Portaluri[a], G. Umbriaco[a,b], V. Viotto[a]

[a]INAF – Osservatorio Astronomico di Padova, Vicolo dell'Osservatorio 5, Padova, 30122, Italy
[b]Università degli Studi di Padova, Vicolo dell'Osservatorio 2, Padova, 30122, Italy

## ABSTRACT

The wavefront sensors used today at the biggest World's telescopes have either a high dynamic range or a high sensitivity, and they are subject to a linear trade off between these two parameters. A new class of wavefront sensors, the Generalised Optical Differentiation Wavefront Sensors, has been devised, in a way not to undergo this linear trade off and to decouple the dynamic range from the sensitivity. This new class of WFSs is based on the light filtering in the focal plane from a dedicated amplitude filter, which is a hybrid between a linear filter, whose physical dimension is related to the dynamic range, and a step in the amplitude, whose size is related to the sensitivity.

We propose here a possible technical implementation of this kind of WFS, making use of a simple holographic diffuser to diffract part of the light in a ring shape around the pin of a pyramid wavefront sensor. In this way, the undiffracted light reaches the pin of the pyramid, contributing to the high sensitivity regime of the WFS, while the diffused light is giving a sort of static modulation of the pyramid, allowing to have some signal even in high turbulence conditions. The holographic diffuser zeroth order efficiency is strictly related to the sensitivity of the WFS, while the diffusing angle of the diffracted light gives the amount of modulation and thus the dynamic range. By properly choosing these two parameters it is possible to build a WFS with high sensitivity and high dynamic range in a static fashion. Introducing dynamic parts in the setup allows to have a set of different diffuser that can be alternated in front of the pyramid, if the change in the seeing conditions requires it.

## 1. INTRODUCTION

The Pyramid WFS [1], in absence of any modulation, is fully equivalent to a Focault test and, in geometrical optics approximation, it allows to retrieve only the sign of the wavefront derivatives. The modulation of the Pyramid WFS distributes the light from the source on all the faces of the pyramid, allowing linear measurements of the wavefront slope. Up to now, Pyramid modulation was mainly achieved using moving parts, like oscillating the prism or using a fast steering mirror [2] [3]. Static modulation was proposed by Ragazzoni [4] and then tested by LeDue using a holographic diffuser [5]. It has also been proposed that residual aberrations while an Adaptive Optics system is running in closed loop may provide modulation [6][7]. However, it may be still advantageous to employ static modulation to enable the Pyramid to better cope with telescope jitter and reduce the manufacturing tolerances on the pyramid apex.

A new class of wavefront sensors, the generalized Optical Differentiation WFS (g-ODWFS), has recently been proposed [8]. The purpose of such a WFS is to break the classical linear trade-off between sensitivity and dynamic range of a WFS, imparting, in its original proposal, spatially varying polarization to the light incident on the focal plane.

In this paper, we propose a possible different implementation of the g-ODWFS, using a Diffractive Optical Elements (DOE) to shape part of the incident light in a ring pattern around the pyramid pin. This provides efficient modulation with the desired radius using a static element, while the remaining part of the light continues un-diffracted, producing a spot on the pin of the pyramid.

## 2. DIFFRACTIVE OPTICAL ELEMENTS

Diffractive optical elements are used to generate complex light patterns with precisely defined dimension in a specific plane. Among the DOEs we analyze here the possibility to use diffractive diffusers to produce a ring of light around the vertex of a glass pyramid, leaving at the same time a given amount of light passing through un-diffracted. Diffusers are far field beam shapers, shaping the beam through interference of a large number of diffractive orders. These elements impart a defined spatial frequency distribution to the phase of the beam and, as the beam propagates, the spatial frequencies in the phase cause the beam to interfere with itself. The resulting beam of a far field optical element is simply the convolution of the Fourier transform of the input beam and the spatial frequencies of the optical element. Diffractive diffusers share many characteristics with diffractive gratings, as they are also far field diffractive optics. In general, a grating is a periodic amplitude and/or phase structure. The resulting electric field emerging from a grating at $z = 0$ is:

$$E_g(x, z = 0) = A(x,0)e^{j(2\pi/\lambda)(n-1)g(x)} = A(x,0)P(x) \qquad [1]$$

Where $n$ is the refraction index of the material constituting the grating, $A(x,0)$ is the amplitude of the input beam and $g(x)$ is a periodic phase function whose height is $\lambda/(n-1)$. In the far field ($z = z'$), Equation (3) becomes:

$$E_g(k_x, z') = F[E_g(x,0)] = F[A(x,0)] * F[P(x)] = A'(k_x, 0) * F[P(x)] \qquad [2]$$

where F represents the Fourier transform, * is the convolution symbol and $k_x = k_0 x'/z$. From Equation (2) we can see that the resulting field at $z = z'$ is simply the convolution of the spatial frequency content of the phase with the amplitude of the input beam after propagating at distance of $z'$. Now, a grating has very distinct orders due to its periodic structure. The spatial frequency composition of the phase is a set of appropriately weighted delta functions spaced at angular intervals regulated by the grating equation:

$$\sin(\theta_t^m) = \sin(\theta_i) + \frac{m\lambda}{\Lambda} \qquad [3]$$

Where $\theta_i$ is the incident angle, $\theta_t^m$ is the transmitted angle of a given diffracted order $m$ and $\Lambda$ is the period of the grating. The far field intensity distribution is then constituted of distinct diffraction orders arranged in the desired spot pattern, but no solid profile is produced (but in the case in which the detector resolution is wider than the actual separation between the discrete diffraction orders). If we now add a second function $h(x)$ with a period much larger than the period of $g(x)$ to the phase in Equation (1), we obtain:

$$E_d(x, z = 0) = A(x,0)e^{j(2\pi/\lambda)(n-1)[g(x)+h(x)]} = A(x,0)P(x)H(x) \qquad [4]$$

And the resulting field after propagating a distance $z'$ becomes:

$$E_d(k_x, z') = F[E_d(x,0)] = A'(k_x, 0) * F[P(x)] * F[H(x)] \qquad [5]$$

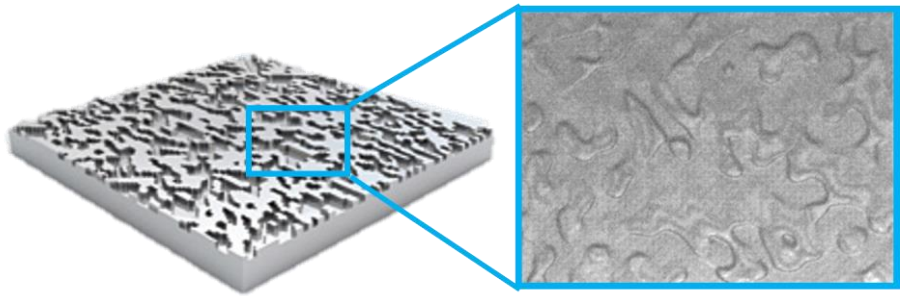

Figure 1: Microscope image of a surface relief phase grating. Levels of gray corresponds to different phase delays induced to the wavefront (and different thickness of the grating).

When this second function is added to the phase, the distinct diffraction orders becomes blurred by the spatial frequency components of the function $h(x)$. Choosing $g(x)$ and $h(x)$ appropriately will produce an apparent continuum of spatial frequencies that results in a solid region filled with light in the far field and we have produced a diffractive diffuser.

Designing and defining properly the diffuser phase functions $g(x)$ and $h(x)$ requires the simulation of an inverted light propagation from the plane in which the pattern has to be created back to the plane of the diffuser. From Equation (5) it is clear that a Fourier transform is involved in this process. The calculation of diffuser designs is absolutely non-trivial, in particular when complex shapes are required, and several algorithm are used nowadays to numerically solve this problem, widely described in other works. We limit here to only hint that the most successful and the most widely used DOE design algorithm is the iterative Fourier transform algorithm (IFTA) [10] for the far field diffraction problems. IFTA was initiated by Gerchberg and Saxton [11] and was designed for the solution of image phase retrieval problems.

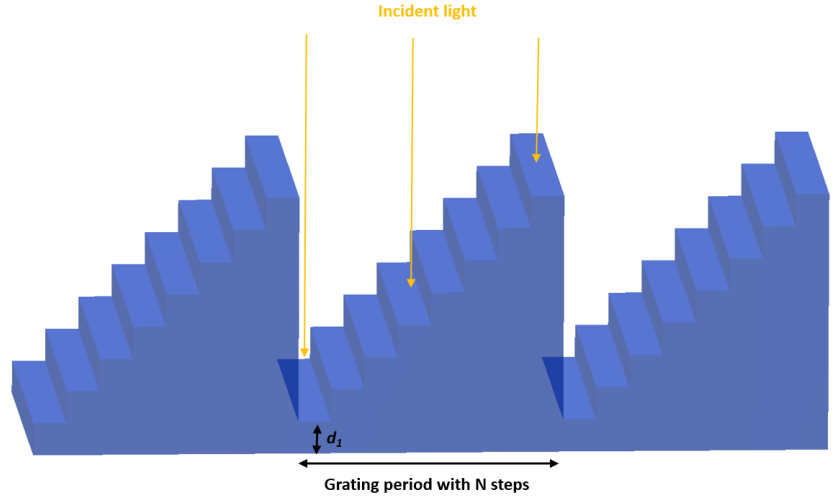

Figure 2: a general periodic diffractive grating.

## 3. MODULATION OF PYRAMID SENSOR

Modulation is fundamental when using a PS for wavefront sensing. In this approach, the aberrated beam is refocused onto the pin of the pyramid, which spatially filters the electric field phasor in the focal plane. Four images of the pupil are then re-imaged on a detector from another lens and their relative intensities give a signal that is directly related to the slope of the wavefront. The signal derived from the re-imaged pupils is given by:

$$S(x,y) = \frac{|I^+(x,y)|^2 - |I^-(x,y)|^2}{|A|^2}$$

[6]

Where $|I^\pm(x,y)|$ are the intensities in the upper and lower pupils respectively and the signal is normalized by the total intensity $|A|^2$. From [9] the Fourier transform of the PS signal is given by:

$$S = \begin{cases} -i\,\text{sgn}(f) & |f| > \frac{\alpha}{\lambda} \\ -\frac{i\lambda}{\alpha} f & |f| < \frac{\alpha}{\lambda} \end{cases}$$

[7]

Where α is the modulation radius and *f* is the spatial frequency. It is thus clear that the PS has two different behaviors: when $f < α/λ$ the signal of the PS in the Fourier domain is proportional to the spatial frequency and it acts as a slope sensor, while in the case of $f > α/λ$ it is sensitive only to the phase. It is clear that the radius of the modulation α determines the linear range where the PS gives a linear response to tilt. In addition, the amplitude of the modulation infers also the sensitivity of the PS, as a small modulation amplitude would produce a large signal even when the aberrations on the wavefront are small, while a large modulation reduces the sensitivity of the PS (but enlarges its linear range).

The modulation of the PS can be performed in several different ways. Originally, it was proposed to vibrate the PS around the focal spot at a rate exceeding the read-out of the detector so that, in geometrical approximation, the rays forming the focal spot will spend some time on each facet of the pyramid, giving a useful signal to compute the wavefront derivatives. The signal of the PS is proportional to the rays deviation and hence to their tilt at the entrance pupil. Another method to implement PS modulation dynamically is to insert a tip-tilt mirror in a plane conjugated to the entrance pupil and then steer the mirror in order to move the focal spot around the pyramid apex, on the four pyramid facets, again at a rate faster than the detector read-out rate.

The same effect can be obtained statically. This possibility was initially proposed and analytically demonstrated by Ragazzoni et al. [4] and then tested and validated, using a 0.5° and 1° diffusers, by LeDue et al. [5]. This kind of holographic diffuser produces a sort of random distribution of the light coming out from the pupil plane, leading to sort of inefficient modulation, as most of the rays are focused in the central region of the light diffused by such device.

We propose here the possibility to use different kinds of Diffractive Optical Elements (DOE), producing a well defined deterministic distribution of the light onto a specifically shaped pattern, and in particular a ring of light to be centered around the pyramid vertex, keeping at the same time a given percentage of un-diffracted light focused on the pin of the pyramid.

The advantage of this approach is clear, as this implementation would allow to have a useful signal from the PS, proportional to the amount of light distributed into the ring around the pin of the pyramid, even in case of large aberrations. On the other hand, the un-diffracted light reaching the pin of the pyramid allows to mantain a high sensitivity also to smaller aberrations, at the same time, similarly to what is proposed for the g-ODWFS.

However, static modulation fixes the sensitivity of the PS, as the radius of the ring around the pyramid vertex is determined by the DOE. This removes one of the advantages of the PS over the Shack-Hartmann Wavefront Sensor, i.e. the capacity to dynamically adjust the sensitivity of the PS by changing the amplitude of the modulation.

Nevertheless, one should consider that the loss in sensitivity with respect to an un-modulated PS depends upon the percentage of the total amount of light that is re-distributed into the ring pattern.

A large fraction of the light diffracted into the ring would result into a high response to large aberrations, but this would damp the signal at smaller scales. Vice versa, if the DOE leaves most of the light un-diffracted, the sensitivity of the PS will be very high to high order aberrations but the response to large aberrations will be much fainter. Thus a deep analysis taking into account the average seeing of the site, $C_n^2$ distribution, required level of correction, adaptive optics techniques to be implemented must be done to find the correct balance into the light distribution on the PS, in order to produce the DOE that better suites at a given telescope site.

Anyhow, it would be also possible, for example, to introduce in the design a motorized wheel accommodating several DOEs with different diffusing angles and different light re-distribution. This allows to restore the variable sensitivity of the pyramid with a layout much simpler than a tip-tilt mirror or an oscillating pyramid.

## 4. CONCEPTUAL LAYOUT

We introduce now a possible implementation of the PS using a DOE for static modulation. The concept is largely based on the probes used in LINC-NIRVANA to acquire NGS for MCAO correction. Each probe there is constituted of a mechanical rail, hosting a couple of lenses magnifying the image of the star on the pin of the pyramid, otherwise too small, and the pyramid itself.

We propose here the same concept, introducing in an intermediate pupil plane a DOE, which produces the already discussed light pattern on the pyramid. Of course the spatial frequencies imparted by the DOE to the phase of the incoming wavefront much be much higher than the spatial frequencies of the modes one wants to measure, in order not to introduce spurious signal on these modes and affect the measurements. It has been demonstrated in [16] that the spatial intensity of the speckle pattern produced by diffractive diffusers is solely determined by the phase changes in the incident beam and are independent from the diffractive diffuser. A display of a possible implementation of such a WFS is shown in Figure 3.

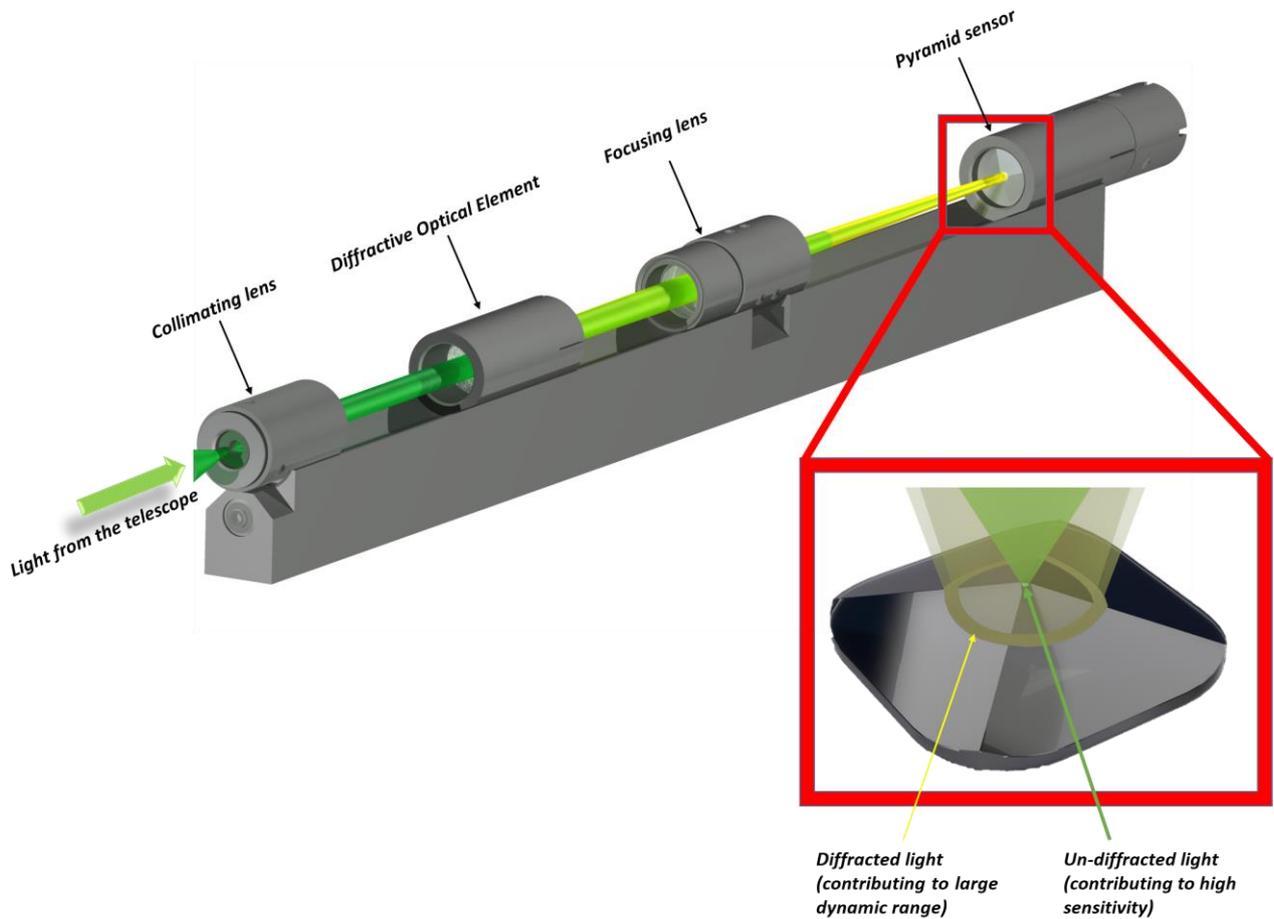

Figure 3: Conceptual idea for implementing pyramid WFS static modulation using a DOE producing a light ring pattern around its vertex, leaving part of the light un-diffracted.

A first lens collimates the light coming from the telescope. This lens is required for two distinct reasons. The first one is that, at the current biggest telescopes the F/# delivered to the instrumentation is quite fast. Positioning a pyramid on the direct focus of the telescope has two main drawbacks. The first one is that the image of a natural guide star on the pin of the pyramid would be very small, which would quickly trigger saturation of the PS. Secondly, such fast F/# entering into the pyramid requires extremely fast optics to re-image the pupils with a reasonable dimension onto a CCD.
The second reason is due to the fact that the DOE must be placed in a plane conjugated to the pupil.
The collimated light passes through the DOE, which impart the desired shape to part of the light of the beam. A second lens focuses the beam onto the pyramid, in the pattern already described. From an opto-mechanical point of view this setup is quite simple to realize and compact, having a single additional optical elements (the DOE) with respect to already existing devices to achieve the goal proposed by the g-ODWFS.

From the design it is clear that the useful signal on the WFS will be result of a composition of the signal due to both the un-diffracted and diffracted light, and the response curve is strongly dependent on the re-distribution of the light from the DOE and the diameter of the modulation ring.

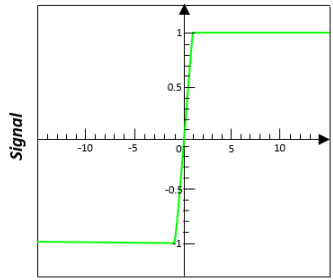

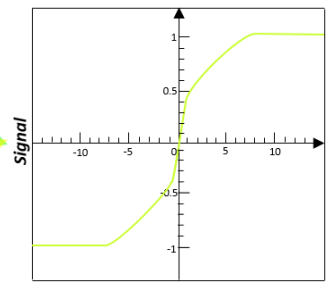

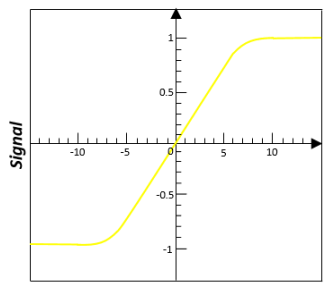

Figure 4: the response of a PS making use of a DOE as discussed in the text is a combination of a not modulated PS with a modulated PS. By sacrificing part of the sensitivity of the not modulated PS it is possible to increase the linear range of the WFS, with a trade-off more advantageous than linear. The loss in sensitivity strongly depends on the amount of light diffracted by the DOE into the ring structure, while the diameter of the ring impose the linear range of the WFS. As an example, in the right panel is shown the response curve of a PS in which 60% of the light is re-distributed into the ring structure, which has an amplitude of $\pm\,6\,\lambda/D$.

It is true that this approach, for its intrinsic static nature, does not allow to tune the amount of modulation to better cope with the seeing conditions. However, it is worth to point out that here the loss in sensitivity is mainly due to the amount of light diffracted to the ring rather than to the amplitude of the modulation, which is present as a second order effect.

In fact, when the seeing conditions allow it, the un-diffracted light provides a high signal, whose intensity with respect to a not modulated pyramid is, in first approximation, reduced of the fraction of diffracted light. On top of this, one should consider that in the high signal regime there is still a contribution from the light in the ring, which is proportional to the ring amplitude.

Moreover, as already demonstrated in [6], the residual of the correction of higher order aberrations acts as a modulation for the lower order aberrations, so the spot at the pin of the pyramid is also modulated, within a certain threshold, by the atmosphere.

By fine-tuning the amplitude of the modulation and the amount of diffracted light one could build a DOE providing good correction performances in a wide range of seeing conditions, even in a static fashion.

While the advantages of this approach upon dynamic modulation (easier implementation) and classical static modulation (lower sensitivity loss to high order aberrations) are clear, we briefly remark here also the main drawbacks of this method, which needs future and detailed investigation.

A first issue is that these kinds of diffusers are usually designed to work with monochromatic light as they exhibit a strong chromaticism. When the diffuser is illuminated with light of different wavelengths from the one it was designed for, the shape of the pattern is preserved but its dimension is not and the light with the longer wavelength will be diffracted on larger angles than the light with the shorter wavelength. The ratio between the dimension of the two images produced on a screen at fixed distance is equal to the ratio of the two wavelengths, as also evident from Equation (3).

Coming back to the PS, this effects translates in a sort of wider modulation for the light with longer wavelength, while the atmospherical turbulence acts exactly in the opposite way, affecting more the light with the shorter wavelengths. In this way we have a sort of different modulations for different wavelengths, with the less aberrated light (toward red) more modulated and the more aberrated light (toward blue) less modulated. The PS will be then more sensitive to aberrations in the blue light, because of the minor modulation, while commonly in adaptive optics a smaller sensitivity and a wider linear range is desired when dealing with large aberrations. In the red part of the spectrum we have the opposite problem, as the sensitivity is lower and also the distortions are smaller. If we introduce then a modulation that is too large there is the possibility to reduce too much the sensitivity of the PS in the red part of the spectrum, accepting thus a large residual from the correction at these wavelength. On the other hand, if we introduce a modulation which is too small there is the risk to saturate the PS in the blue part. Of course these are the worst cases, and a lot of modulations in the between among which choose are possible. Having clear in mind that the perfect modulation with such a chromaticism does not exist, because when optimizing the residuals of the correction on one side, on the other side the residuals become poorer, the correct modulation is a trade-off between the optimal solutions derived for the boundary wavelengths used for the correction.

A second, and equally important, effect of the chromaticism is the wavelength dependence of the diffraction efficiency, which represents the percentage of the incident light redirected by the desired angle. In particular we are interested in understanding which is the diffraction efficiency of the zeroth order, i.e. the light not diffracted. Calculating the diffraction efficiency of a generic diffuser is not trivial and requires electromagnetic theory to find rigorous numerical solutions to the full vector field equations. However, when the size of the diffracting feature on the DOE is large compared with the wavelength of the incident light (typically at least 5 times larger) a simplified model can be used to predict diffraction efficiency of a DOE. In this case we can use the scalar approximation to Maxwell's equations, in the assumption that light can be treated as a scalar rather than vector field and that electric and magnetic field components are decoupled. According to the scalar theory, the diffraction efficiency of an arbitrary *l*th order produced by a periodic diffractive grating with sub-period N in case of normal incidence of the light on the grating is:

$$\eta_l^N = \left[ \frac{sin\left(\pi \frac{l}{N}\right)}{\pi \frac{l}{N}} \frac{sin\left[N\left(\frac{1}{2}\varphi_1(\lambda) - \pi \frac{l}{N}\right)\right]}{Nsin\left(\frac{1}{2}\varphi_1(\lambda) - \pi \frac{l}{N}\right)} \right]^2$$

[8]

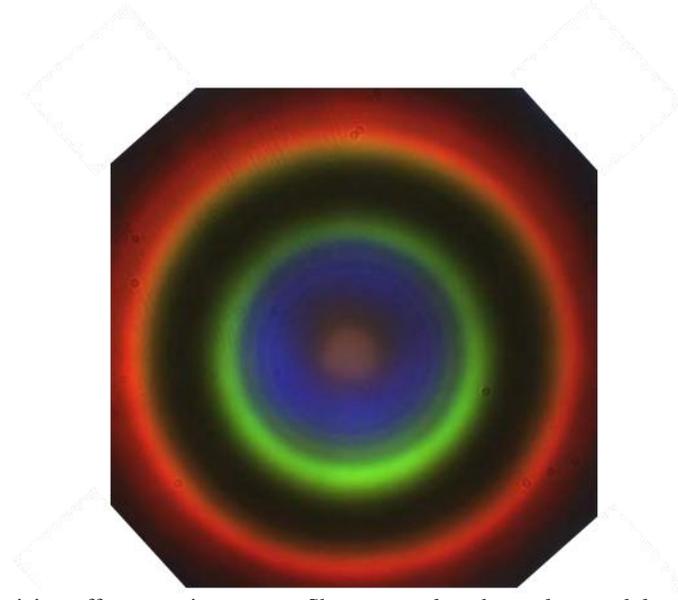

Figure 5: chromaticism effect on a ring pattern. Shorter wavelengths are less modulated that longer ones.

Where $\varphi_1(\lambda) = 2\pi\ [n(\lambda)-1]d_1/\lambda$, $d_1 = \lambda_0/(n-1)N$ is the geometrical height of the step of the grating, $n(\lambda)$ is the refraction index of the step material and $\lambda_0$ is the design wavelength. Within the scalar theory region of validity, Equation 8 can determine the amount of light in any diffraction order for any number of phase levels. It is clear from this equation that the diffraction efficiency varies as a function of the wavelength, and once the grating is optimized to work at a certain design wavelength $\lambda_0$ its diffraction efficiency will drop off at different wavelengths. Thus, when operating at wavelengths far from the design one, the zeroth order will be enforced, and in general all the light will be distributed at different percentages in the diffraction orders.

It has been demonstrated that in the scalar theory, the diffraction efficiency of an arbitrary DOE can be directly related to the diffraction efficiency of a grating [12], so Equation 8 can be generalized to any arbitrary phase diffractive structure.

Now, if we assume we want to produce a DOE producing a ring between 0.15° and 0.3° at a wavelength of 632.8 *nm*, the smallest spatial frequency expected in the diffuser phase profile should be of the order of:

$$d = \frac{\lambda}{\sin(\theta_{max})} \qquad [9]$$

Corresponding to a minimum feature size, for the case under evaluation, of 120 *μm*. Since the wavelength tipically used for adaptive optics correction using natural guide stars is in the band 0.4 – 0.9 *μm*, we are well inside scalar theory regime and so we can use the equations presented here to describe such a diffuser.

However, DOE chromaticism can be reduced, even if not cancelled, coupling it with a refractive element that negates its chromaticism. Other approaches, like using multi-materials with different refraction index to produce the DOE, have been proposed. [13] [15]

By aligning two DOEs made of different dispersive materials, the optical path differences as a function of the wavelength can be controlled. Applying this concept, a combined achromatic DOE can be designed for applications that require wavefront control using different wavelengths or wide band light sources. [14]

A second issue could arise directly from the diffractive nature of the DOE. As we have seen before, in order to produce diffraction, the DOE must have features which are on the scale of the wavelength of the incident light. These are, most commonly, etchings in the substrate of the diffuser. The DOE makes the pencil of rays that pass through it to fan out, and acting like a light source. However, since the DOE is part of the pyramid sensor, it is optically conjugated to the detector, and an image of its surface will be seen on the CCD. Thus, in this configuration, the brightness variations on the pupil images are unavoidable.

This is not a problem for the wavefront analysis, as this pattern can be recorded during the calibration phase of the PS, but rather for the dynamic range of the CCD itself, which must accommodate these brightness variation across the pupil. The result is that it is not possible to use the full dynamic range of the CCD for the slopes measurements.

Other issues can be related to the manufacturing of the DOE itself and to the necessity to not introduce spurious perturbations to the modes that are intended to be measured. Moreover, even if a continuous phase profile would be ideal for this application, the DOE are usually fabricated with discrete levels. These discrete levels cause higher order stray light in the system, which may produce other unwanted orders in the output pattern and thus a loss of useful light on the PS, which is crucial in adaptive optics. Even though this effect can be minimized in the design phase of the DOE, fabrication errors (even of few nanometers) usually manifest into an increase of the zero order light and, generally, into a different distribution of the light among the diffraction orders.

This requires a careful design analysis of the problem and a detailed design of the DOE, which requires the use of complex algorithms, together with very accurate fabrication techniques.

## 5. CONCLUSIONS

Summarizing, diffractive optical elements can be theoretically used to reshape the white light from a natural guide stars in a ring pattern around the vertex of a PS to provide static modulation, while keeping part of the light on the pin of the pyramid, providing high sensitivity. The advantage of these diffusers is that the technology to produce them is widely diffused, as it is basically the same used to print integrated circuits (photolithography). The most problematic aspect to face is the strong chromaticism affecting the DOEs. This would result in a multi-ring pattern when illuminating them with white light (as in adaptive optics) and then in different modulations of the light with different wavelengths. However, at the current status of technology, DOEs can be produced coupling materials with different refraction indices, or coupled to refractive elements negating the DOE chromaticism, to provide nearly achromatic characteristics on a wide spectral band, making these diffusers attractive for PS static modulation future studies.

A possible opto-mechanical layout has been proposed here, which would be of simple realization and with a minimal amount of optical elements to achieve a large dynamic range maintaining a high sensitivity to high order aberrations. Despite being in principle a very promising technique and of easy mechanical implementation, the drawbacks briefly described in this paper must be accounted and investigated with a detailed analysis, to verify how they exactly affect the performance of a PS.